\documentclass[12pt]{article}
\usepackage{graphicx}
\begin{document} 
\baselinestretch
\renewcommand{\baselinestretch}{1.5}

\noindent
\textbf{\large
{On the relation between structural diversity\\
and geographical distance among languages:\\
observations and computer simulations}}

\bigskip
\noindent
Eric W. Holman, Christian Schulze, Dietrich Stauffer \& S{\o}ren Wichmann

\bigskip
\textbf{Abstract}

\small
Since the groundbreaking work of Nichols (1992) it has been clear that the use of typological databases for making inferences regarding linguistic prehistory could potentially have much to offer. The recent availability of larger typological databases such as Haspelmath et al. (2005) has brought the linguistics community closer to having a solid, empirical foundation for making actual claims about prehistoric migrations, deep genealogical relationship, and patterns of areal linguistic interaction. Nevertheless, still more data are needed, and more than anything a number of methodological problems need to be adressed. These problems, which are the focus of this paper, include the following. How might we go about distinguishing diffusion from genealogical inheritance when looking at structural similarities among languages? For how long may we expect to continue to see a difference between traces of relatedness and traces of diffusion? What effects do factors such as speed of migration, the time depth of interaction among a certain set of languages or the rate of diffusion have on the similarities among initially related languages on the one hand and initially unrelated languages on the other? These questions will be addressed from two perspectives. The first perspective is an empirical one, where observations primarily derive from analyses of the data of Haspelmath et al. (2005). The second perspective is a computational one, where simulations are drawn upon to test the effects of different parameters on the development of structural linguistic diversity. The results suggest that there is indeed some hope that we may derive new empirical insights regarding linguistic prehistory by drawing upon typological data. We do not, however, make any specific empirical claims in this paper, but instead concentrate our efforts on methodology.

\bigskip
\noindent
\textbf{Keywords}\\
Spatial autocorrelation, computer simulation, isolation by distance, areal typology
%newtext: "Spatial autocorrelation" is now used instead of "sptial correlation". It is a standard term and means "the correlation of a variable with itself through space". It is an important keyword even if we don't actually use the exact term in the text, but rather paraphrase it.

\normalsize
\section{Introduction}

This paper investigates the relationship between typological similarity and geographical distance among languages. Because of the effect of diffusion (transfer of structural features among languages due to contact among speakers) it is expected that we should observe the phenomenon known among population geneticists as \small{ISOLATION BY DISTANCE} (IBD), that is, a relation where increased geographical distance results in greater differences\textemdash in this case among languages. This relationship is expected to obtain whether or not the languages in question are related. Nevertheless, while an IBD-effect is expected to occur universally, the effect might be enhanced or diminished by different factors such as initial similarity among language (henceforth domination) vs. initial non-similarity (henceforth fragmentation), rates of diffusion, rates of internal language change, speed of migration, and perhaps other factors. How do these various factors affect the relationship between structural similarity and geographical distance? For instance, should we expect that languages which are related tend to be as different from one another as languages that are not related at great geographical distances? Should we expect that the relationship between structural differences and geographical distance among the languages in a family tend not to be distinguishable from the relationship between structural differences and geographical distance among languages that are not related if, for whatever reason, the former languages have suffered more diffusion, migration or internal change?

Another set of questions which this paper will be concerned with is to what extent qualitatively different typological datasets may have an effect on our ability to discern the footprints of language history. More specifically, we shall  investigate whether it makes a difference when one uses binarily as opposed to ternarily or quarternarily etc. encoded features and we shall also look at whether or not it has an effect on one's results whether the values of the features used stand in a gradual relationship to one another or not (illustrative examples and more clarification of this issue will be given below). 

The problem area will be introduced by means of empirical examples produced by comparing structural diversity among languages of Eurasia, Africa, and the New World as a function of geographical distance. These examples, which merely serve as an illustration of the types of differences in distributions that the typologist is likely to encounter, are introduced in the following section. Because of limitations in the data the examples can merely serve as illustrations of how actual empirical distributions \textit{might} look like; they do not suffice to serve as actual case stories of how aspects of prehistory may be inferred from linguistic data. The remainder of the paper presents various computer simulations from which distributions which are in some respects similar to the empirical examples are be predicted to occur. By varying the parameter settings we investigate the two sets of questions raised in the above paragraphs. 

To our knowledge, computer simulations have never before been used to answer the specific questions raised here regarding the relationship between structural differences and geographical distances, nor have they been used to investigate the properties of typological datasets. Nevertheless, a small but growing number of researchers have drawn upon computer simulations and mathematical models to investigate other aspects of linguistic evolution, including the development of linguistic diversity (Abrams and Strogatz 2003, Sutherland 2003, Patriarca and Lepp\"anen 2004, Mira and Paredes 2005, Schulze and Stauffer 2005, Wang and Minnett 2005, Kosmidis et al. 2005, Schw\"ammle 2005, Oliveira et al. 2006a-b, Pinasco and Romanelli 2006), the development of taxonomic dynamics (Wichmann et al. 2006), language change (Nettle 1999a, Niyogi 2002, 2004; Prevost 2003, Baxter et al. 2006), and the evolution of language structure (Cangelosi and Parisi 2002, Nowak et al. 2002, Christiansen and Kirby 2003, Wang et al. 2004, de Boer 2006, Niyogi 2006). As regards the investigation of empirical correlations between linguistic differences and geographical distance there are precursors in the field of dialectometry, which was initiated by S\'eguy (1973) and developed further in many subsequent works, including Goebl (1984, 2005), Nerbonne et al. (1996), and Kretzschmar (1996).
%newtext I put in Goebl 1984 instead of Goebl 1982, because the former is his main work; I excluded "See Goebl (2005) for an overview" since this mostly deals with his own work

\section{Empirical examples: distributions of language diversity in Africa, Eurasia, and the New World}

The data drawn upon in this section are provided by \textit{The World Atlas of Language Structures} (Haspelmath et al. 2005, henceforth WALS).  WALS contains 138 maps showing the distribution of different phonological, lexical, and grammatical features for a sample of languages that varies in size among maps from roughly 100 to 1200 languages.  The present study draws on 134 of the 138 features, excluding features that involve redundant data.  Each feature has anywhere from two to nine discrete values.  The total number of languages from which data are drawn in WALS is 2560.  The present study excludes pidgins, creoles, and sign languages, leaving 2488 languages.  The classification used in WALS, which we also adopt here, represents an attempt to follow the views taken by the majority of specialists and results in 205 families and isolates.  The families are defined on the basis of inheritance, and languages in different families have no generally acknowledged common ancestor.  In order to investigate diffusion as distinct from inheritance, the present study includes all possible pairwise comparisons of languages in different families, and no comparisons of languages in the same family.

The difference between languages in different families as a function of the geographical distance between the languages was measured in the following way. For each pair of languages, their distance was calculated from the latitudes and longitudes in the WALS database, where the location of each language is defined as a spot somewhere near the center of the region where the language is spoken (see Comrie et al. 2005: 7 for more detail). Pairs of languages were then grouped according to distance in ranges such as 0-500 km, 500-1000 km, 1000-2000 km, etc. For each of the 134 features, the average difference between the paired languages in a group was defined as the number of pairs with different values of the feature, divided by the number of pairs for which feature was attested in both languages. These proportions were averaged across the 134 features and expressed as a percentage to represent the overall difference of the language pairs in a group. Figure 1 plots difference as a function of mean distance, separately for languages in Africa, the New World, and Eurasia.

\begin{figure}[h]
\begin{center}
\includegraphics[scale=0.5,angle=-90]{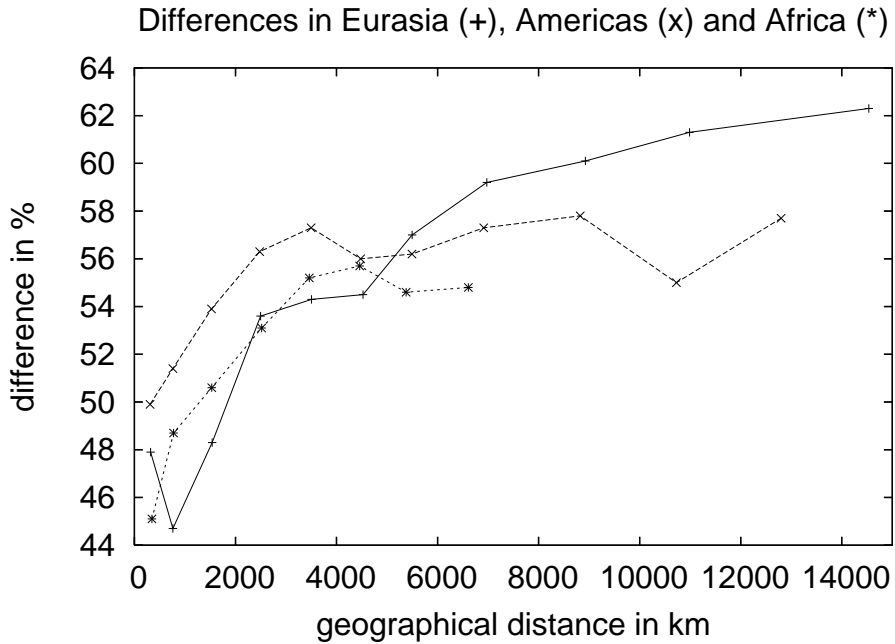}
\end{center}
\caption{The relationship between geographical distance in km and typological similarity in Africa, the New World, and Eurasia
}
\end{figure}

All three curves show a clear effect of diffusion.  Difference is least between languages less than 1000 km apart, and then increases with increasing distance.  Superimposed on this general trend are some possible differences among the curves.  For languages less than about 5\,000 km apart, the curve for the New World suggests greater differences between languages than do the other curves.  For languages more than about 5\,000 km apart, the curves for Africa and the New World appear to approach asymptotes, while the curve for Eurasia continues to rise, but increasingly less so the larger the distance. Despite differences, all three curves show the effects of isolation by distance (IBD). This term, which we borrow from population genetics, was defined by Wright (1943) and refers to situations where large genetic correlations are found among spatially proximal populations and drop off smoothly as the distances among the populations decrease. Discussion and examples are found in Epperson (2003:14-25). One illustrative example is the study by Sokal and Menozzi (1982) of different allele frequencies for HLA blood group loci in European and Middle East populations. Large autocorrelations were found within a range of approximately 700 km, lesser positive values within an approximate 700-1\,400 km range and negative values beyond this range. %newtext
A notion from geography which is similar to IBD in biology is `\small{TOBLER'S FIRST LAW OF GEOGRAPHY}, which states that ``everything is related to everything else, but near things are more related than distant things'' (Tobler 1970). %endnewtext
For linguistics, Nerbonne and Kleiweg (forthcoming) define what they call `\small{THE FUNDAMENTAL DIALECTOLOGICAL POSTULATE}' as follows: ``Geographically proximate varieties tend to be more similar than distant ones''. Their Figure 1, which is very similar to our Figure 1, illustrates how differences among dialects depends on distance (similar curves have been published by S\'eguy 1973 and Goebl 1984). This principle or postulate, then, is equivalent to IBD, but we shall use the latter term because it has a longer history of use.
%newtext: the parenthesis with references to Goebl and Seguy is new

It is appropriate at this point to introduce some hedges about the differences among the three curves.  For conventional statistical tests, the units of analysis must be independent of each other.  Language families are obviously not independent with respect to diffusion between them; moreover, the 134 features are not independent either, because some are known to be related to each other while others may also be related even if the implicational relations have not yet been discovered by typologists.  Consequently, it is not clear whether the relatively small percentages that separate the curves are statistically significant.  We must therefore restrict ourselves to presenting these examples as possible examples of different relationships between structural difference and geographical distance. As we shall see, some support for the hypothesis that they are, indeed, possible examples is provided by computer simulations. Taken together, the suggestive empirical data and the simulations about to be presented provide sufficient motivation for going on to consider various ways to explain such presumably possible distributions. Whether Figure 1 adequately reflects differences among languages in different parts of the world is a question which cannot be approached until more language data and a statistically better supported understanding of typological dependencies become available. This may well take several years.

\section{Differences in IBD-effects and their possible causes}

\subsection{Intuitive/common-sensical explanations}

Looking at the examples in Figure 1 and assuming, for the sake of the argument, that it adequately reflects differences among these three parts of the world, how might we then explain such differences? The following are some factors that intuitively could be involved in producing the differences in the curves.

\textit{Differences in diversity.} 
If the Americas were populated by a single wave of migration, as a growing body of genetic evidence suggests (most recently Stone and Stoneking 1998, Silva Jr. et al. 2002, Tarazona-Santos and Santos 2002, Zegura et al. 2004) then all New World languages might share a single ancestor or descend from relatively similar languages. At the time when the migration would have taken place, the Eurasian languages would already have suffered thousands of years of differentiation. If such a relatively recent shared ancestry of New World languages obtains, it seems likely that these languages, even at great geographical distances, could be more similar than Old World languages spoken similarly large distances apart. This whole scenario, however, seems to be belied by the African languages, which, at any distance, are more similar than the New World languages. On the other hand, there might also be an effect pulling in the opposite direction of less diversity which starts to set in when a continent is fully populated (cf. Nettle 1999b).

\textit{Rate of migration.} Leaving aside Africa for the moment, we could compare the curves for the New World and Eurasia. We do not know how long it took until the Americas were populated. The archaeological evidence tells conflicting histories. On the one hand, the migration of paleo-indians is supposed to have involved Beringia, on the other hand the earliest archaeological evidence for human presence comes from caves in South America. Nevertheless, it seems likely that the migration went relatively fast. A fast migration into a territory which was presumably previously not occupied might account for the difference at large distances between the curves for New World and Eurasia.

More similarity at short distances in the Old World areas of Africa and Eurasia might be a symptom of sedentism. When farming is the prevalent subsistence pattern people tend to migrate less. At short distances people belonging to different ethnic groups would therefore enter into prolonged contact. Inversely, if the dominating means of subsistence is hunter-gathering, the same geographical places would not witness contact taking place among certain ethnic groups for very long periods at a time. Different groups would acquire new neighbors at relatively short intervals. Therefore, at short distances languages would be less similar in hunter-gatherer societies than in farmer societies. If this hypothesis is correct and if it is correct that there are significantly more hunter-gatherers in the New World than in Eurasia and Africa, then we have an explanation for the differences in the three curves at 0-5\,000 km.

\textit{Rate of diffusion.} Possibly languages in the Americas tend to show less diffusion (i.e., less transfer of features from one language to the other) than languages in the Old World, for one or the other reason. We could imagine, for instance, that the group(s) who were successful in initially populating the New World were endogamous. If there was just one group initially it would actually have had to be endogamous because there would not be any other group from which to pull marriage partners. Endogamy could have continued to be a prevailing trait among Native Americas for enough time that an effect of more differences among New World languages at short distances (0-5\,000 km) compared to areas of the Old World might obtain (this argument is due to Cecil H. Brown, in personal communication, 2006).

These various thought experiments lead us to doubt that there will be simple and monolithic explanations for differences among linguistic IBD-effects. Instead, they suggest that a variety of factors could affect the types of distributions that we have been looking at. Factors that pull in the same direction may be difficult to tease apart, and it may be difficult to judge what the relative impacts of factors that pull in opposite directions are. It is therefore doubtful that at present we are going to make any progress in the study of linguistic IBD-effects from initially analyzing empirical data, even if the ultimate goal is to understand real-world situations similar in kind to the ones that are vaguely reflected in the data producing Figure 1. We instead turn to computer simulations which will allow us to look at different factors that might possibly affect linguistic IBD-distributions in isolation from one another. The following subsection presents a computational model suitable for such an investigation and subsequently we present the results of the implementation of this model.

\subsection{Computational model for testing different hypotheses}

In previous studies applying computer simulations to problems of language evolution different models have been developed. The `\small{VIVANE MODEL}' (Oliveira et al 2006a-b) simulates the first occupation of a large continent by human beings who initially all speak one language, and the growing diversity of languages during this colonization. However, languages in that model were simply numbered consecutively, preventing a simulation of structural differences shown empirically in Figure 1. The language learning model of Nowak et al (2002) has a similar disadvantage. The computer models of Abrams and Strogatz (2003) (followed by Patriarca and Lepp\"annen 2004 and Pinasco and Romanelli 2006), of Kosmidis et al. (2005), and of Schw\"ammle (2005) deal with relatively few languages, not with the thousands reflected in Figure 1. Thus we modify the `\small{SCHULZE MODEL}' (Schulze and Stauffer 2005, 2006) (also used by Te\c sileanu and Meyer-Ortmanns 2006). Since readers of the present paper may not know this model or may not have easy access to the physics literature we describe it in some detail in the following.

A large square lattice is occupied by people speaking one language each.  Each language (or grammar) is characterized by $F$ features, each of which can take $Q$ different integer values from 1 to $Q$. Mostly we use $F=8$. For the simplest binary case $Q$ efficient bit-string algorithms have been used in the past, allowing larger $F$, but for the present purposes we vary $Q$ up to 9 and thus use only simpler programs (written in Fortran and available as langpotts26.f from stauffer@thp.uni-koeln.de). 

Initially, only the top line of the lattice is occupied, and all others are empty. The people in the top line either all speak the same language (``\small{dominant}'' start) or each one independently selects randomly one of the  $Q^F$  possible languages (``\small{fragmented}'' start). Mainly, what we are interested in is to study the differences between fragmented and dominant initialization. (For dominance, the one initial language has the integer part of $(1+Q)/2$ for all its 
features, i.e. the central value for odd $Q$). 

For each time step (human generation), each occupied lattice site $i$ can change its features and that of its neighbours following four probabilities $p,q,r,s$ for four different processes i to iv:

\bigskip
i) Shift ($r$): If a fraction $x$ of people in the whole population speaks the language of site $i$, then site $i$ shifts with probability $(1-x)^2r$ to the language of one of its four lattice neighbours, randomly selected. (If this site is still empty, the new language is that used for the initialization, i.e. either the dominant one or a randomly selected one.) This shift takes into account the tendency of humans to give up speaking minority languages.
 
ii) Change ($p$): Each of the $F$ features is randomly changed with probability $p$. For zero diffusion probability (see next process) this change is to a randomly selected value for unordered features and to the old value $\pm 1$ for ordered features; see section 4 for this distinction. (If in the ordered case the new value would be 0 or $Q+1$, the old value is kept for this feature.) This change describes the language changes from one generation to the next.

iii) Diffusion ($q$): In the case of diffusion, with probability $q$ the new value is taken is taken during the change of process ii from one of the four lattice neighbours, selected randomly and independently for each of the $F$ features. In this way it is simulated how a language may take over traits from other languages. (If this neighboring site is still empty no diffusion takes place.)

iv) Migration ($s$): Each of the four nearest neighbours (North, East, South and West) is checked, and nothing happens if it is already occupied. If it is empty, then with probability $s$ it becomes occupied, with the same language features as on the original site $i$. This original site $i$ remains occupied. Migration, then, simulates the peaceful colonization of uninhabited territory by an expansion of the population.

\bigskip
In earlier papers published in physics journals, the first three processes were respectively denoted by the terms `flight', `mutation', and `transfer', while process iv, which was not introduced in earlier papers, presumably would be called `diffusion' there. 

The differences between languages on the top line and those on the lattice line separated by $d$ lattice spacings are calculated in two different ways for the ordered and the unordered features, discussed in Section 4 below: as the average number of features which are different (unordered case), and as the average sum of the absolute differences in the features (ordered case). For binary features, $Q=2$, this distinction vanishes. (In both cases we average only over occupied sites.)

The Schulze model and its variants were simulated without migration in several publications (mostly reviewed in Schulze and Stauffer 2006), as a function of the three probabilities $p,q,r$ and the total population $N$. For $r \simeq 1$ and large but finite $N$ a sharp phase transition was found between fragmentation and dominance: Either the system ends up fragmented, when each possible language is spoken by about the same number of people (if the population is not large enough, then a roughly random selection of all possible languages is spoken). Otherwise the system ends up dominated by one language spoken by the majority while the others mostly speak minor variants of this dominating language. Both final states, fragmentation or dominance, can be reached either from random fragmentation or from dominance. If we start with one person whose offspring lets the population grow up to a final stationary value, we necessarily start with dominance, and if we also end with dominance, we may have a maximum of the number of spoken languages at some intermediate time (Schulze and Stauffer 2005) as in Nettle (1999b). In the $p-q-$plane, one finds a transition line separating fragmentation (large $p$, small $q$) from dominance (small $p$, large $q$). If instead we select $r \ll 1$, final dominance may become impossible. In the present simulation of $L \times L$ square lattices, the population is $N = L2$ and should be compared with the possible number $Q^F$ of languages. Perhaps for $N \rightarrow \infty$ also the time which dominance needs to emerge from fragmentation goes to infinity; thus mathematical limits should be considered with caution. 

\subsection{Results of the implementation of the model}

\begin{figure}[h]
\includegraphics[scale=0.5,angle=-90]{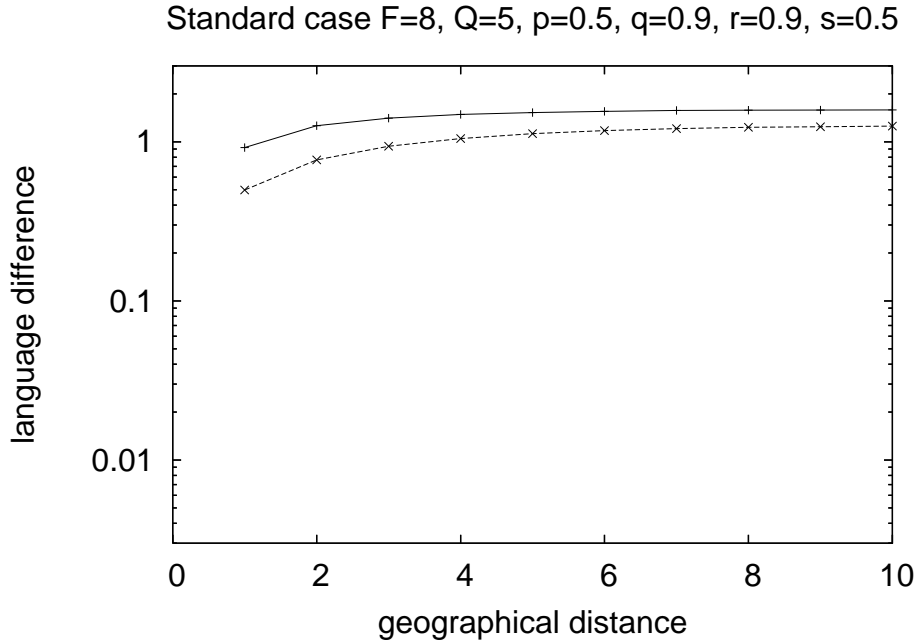}
\caption{A simulation using standard parameter settings. Legend: + = fragmentation, x = dominance.}
\end{figure}

A simulation using standard parameter settings is shown in Figure 2. Here and in the following figures we show the result after 70 iterations, when the results no longer change systematically. Only the topmost 10 lines of the $10001 \times 100001$ lattice are analyzed, comparing pairs of sites with the first one in the top line and the second one at distance $d$ exactly below the first. (The values remain constant also for longer distances up to 30.) We start from fragmentation (+) and dominance (x). (For fragmentation, already after 5 iterations the results no longer change.) For larger diffusion $q\simeq 1$ instead of 0.9, the initially fragmented population would have changed later into one dominated by only one language, cf. Schulze and Stauffer (2006) and Stauffer et al. (2006). We assume that features are ordered and measure differences between languages accordingly (see Section 4 below for more detail).

The standard parameter settings, then, are: \\
\\
\indent internal language change: $p=0.5$ \\
\indent diffusion: $q=0.9$ \\
\indent language shift $r=0.9$ \\
\indent migration: $d=0.5$ \\
\indent number of features $F=8$ \\
\indent number of feature values (states, choices) $Q=5$ \\
\\
Since at present we do not know how to translate empirical data into expected absolute probabilities for language change, diffusion, language shift and migration, all the values that we operate with should be looked upon as highly abstract. The same goes for the results of measuring structural differences and geographical distances. Nevertheless, a comparison of the empirical data with the results of the computer simulations suggests that a simulated distance of 2 roughly corresponds to a real-world distance of 5\,000 km. As suggested by Figure 1 as well as an average over all of the world's languages (not shown here) this is the distance at which there tends to cease to be a significant inverse correlation between distance and structural similarity among languages of different families. With one exception (see Figure 4) all the simulations for the corresponding distance of $\approx$2 similarly show a weakening of the correlation.

An important result of the simulations shown in Figure 2 is that a difference between initial fragmentation and initial domination continues to be preserved over long distances, even if this difference is diminished somewhat. This would mean that given two situations where all else is equal, we may be able to distinguish between languages sharing a common ancestor and unrelated language by means of typological data. As we shall see, this `preservation of history' does not result from all parameter settings, but it is the rule rather than the exception. 

In the following we shall vary the settings to study the effects of each individual parameter.

\subsubsection{Rate of diffusion}

\begin{figure}[h]
\includegraphics[scale=0.5,angle=-90]{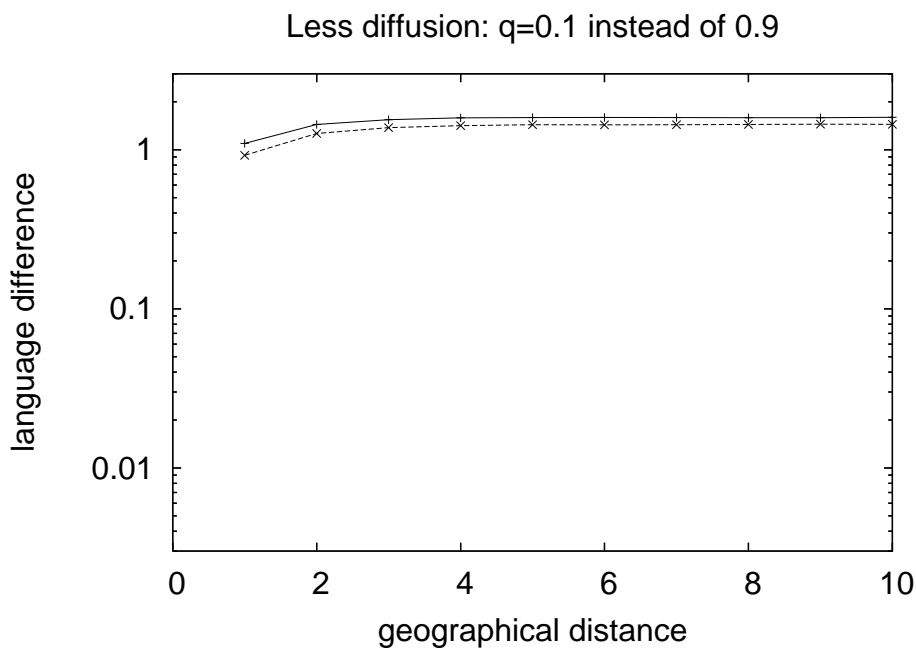}
\caption{The effect of varying diffusion rate from 0.9 (Figure 2) to 0.1 (this figure). Legend: + = fragmentation, x = dominance.}
\end{figure}

Decreasing the rate of diffusion makes initially related language more different, but the effect of such a decrease diminishes with the distance. There is hardly any effect to be seen for fragmentation, cf. Figure 3. That is, diffusion has a greater effect on the degree of similarities among related than among unrelated languages. This is a highly interesting result which is initially somewhat counter-intuitive, but it can nevertheless be brought to accord with the real-world situation. We know that diffusion is a highly potent force in language change. Nevertheless, we also see an immense structural diversity among the world's languages. What could explain this apparent discrepancy is that whereas diffusion may make languages that are in contact with one another more similar, these regional similarities contribute to inter-regional diversity. It \textit{is} possible to change from initial fragmentation to domination, but that requires the diffusion rate to be 0.999, as witnessed by simulations not shown here.

The structural diversity of initially similar languages is apparently more sensitive to the degree of diffusion than that of initially dissimilar languages. This is because when we start with fragmentation, we start with a random distribution and also end up with a nearly random distribution. When we start from dominance we start from a completely ordered distribution and induce changes leading to a nearly random distribution. That involves a greater difference in the distribution. 

\subsubsection{Rate of migration}

We have tried to vary the migration probability to 0.1 and let the simulation run for 200 iterations to get stationary results. This has is virtually no effect. The curves are so similar that no difference can be made out visually although there are minor differences in the data. Possibly this lack of an effect relates to the set-up of the simulation where only members of fully-occupied lattice lines are compared. 

\subsubsection{Rate of language shift}

A simulation where the rate of language shift was varied from 0.9 to 0.5 showed that this has no effect in the case of initial fragmentation. Although more language shift should reduce the number of languages, it does not affect the overall structural diversity as measured in the total number of differences among language pairs. It stands to reason that, in the extreme case where the world's population was divided up into, say, speakers of Chinese and English, the differences among these two languages might still correspond to the average differences among the current languages of the world. The rate of language shift does seem to have a small effect in the case of dominance. With more shift, the offspring of an initially uniform language become more similar, but this effect evaporates at large distances. Again, this is intuitively obvious. As in the case of the standard parameter settings the curves continue to be distinct at large distances. Given the high degree of similarity with Figure 2 we do not show the graph here.

\subsubsection{Rate of language change}

\begin{figure}[h]
\includegraphics[scale=0.5,angle=-90]{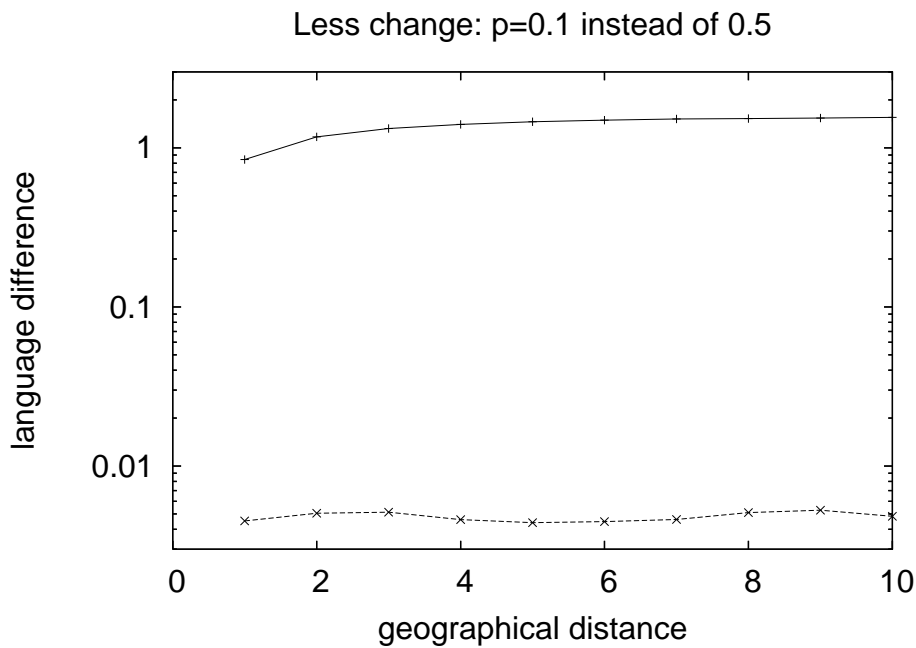}
\caption{The effect of varying the rate of language change from 0.5 (Figure 2) to 0.1 (this figure). Legend: + = fragmentation, x = dominance.}
\end{figure}

Figure 4 shows that reducing the rate of language change has no effect when initial fragmentation obtains, but has a drastic effect in the case of initial domination. In the former situation the effects of language changes, whether there are many or few, will tend to cancel one another out. In the latter situation internal language change is a major contributor to diversity. We have seen in 3.3.1 that the diffusion rate also affects initially similar languages, but only a small effect was found although we varied the probability from a low of $q=0.1$ to a high of $q=0.9$. Again, the real-world situation supports these results, which initially are perhaps surprising: given that there is a lot of structural diversity among the world's languages, internal change $has$ to be a strong factor since it has apparently continued to outweigh the combined effects of language change and diffusion from the beginning of the evolution of language several tens of thousand years ago to this very day. (This is not to say that we may forever continue to see the amount of diversity that we see today.)

\subsubsection{The transitional effect of a fully populated lattice}

\begin{figure}[h]
\includegraphics[scale=0.4,angle=-90]{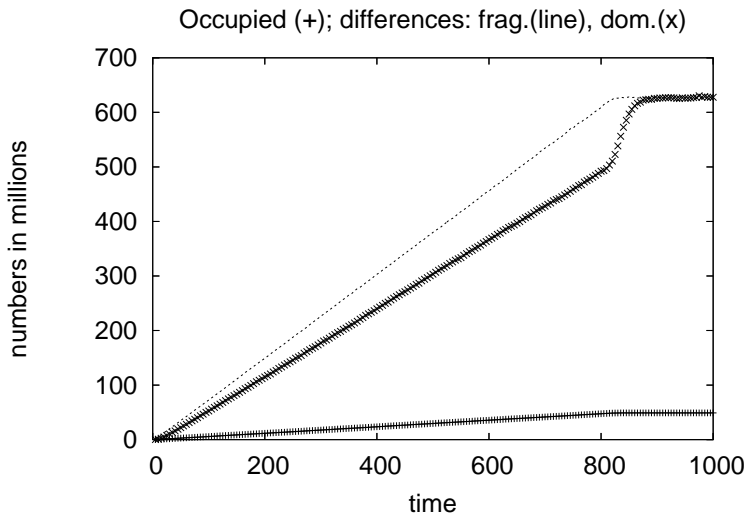}
\caption{Results of a simulation showing the behavior up until the total $7001^2$ or 49 million sites of the lattice are occupied and continuing for some 200 time steps beyond (yet more iterations did not result in further changes). The lower curve shows the number of occupied sites, the upper two curves the sum over all differences, summed over all occupied sites. The standard parameter settings, described in section 3.3, were used.}
\end{figure}

Up to this point we have seen that a difference between initial domination and fragmentation is preserved for large geographical distances. In Figure 5 we show what happens at many time steps when the lattice becomes fully populated. This is the point where the lowermost curve, which indicates the number of occupied sites, no longer increases. Initially, as we have also seen in the previous graphs, both fragmented and dominant start go asymptotically towards infinity; here this effect is shown by a line in the plot because we sum over the differences among the languages rather than showing averages. Then, at the transition point of complete occupation the initially different languages immediately reach a maximum of differences. At the same point, the initially similar languages rapidly increase their differences and soon reach the same maximum as the initially different languages. From there on, nothing new happens: the systems have entered into equilibria. From this we can conclude that the overall difference between initial fragmentation and initial domination ceases to become preserved when all people are surrounded by neighbours with which to interact.

A plot of the differences by distance as done in the previous figures indicates that for $t=1\,500$, when the whole lattice is occupied, the curves for initial fragmentation and initial domination agree, as would be expected. There is the usual IBD-effect, producing a rise in differences up to a distance of 2 and continuing but flattening out beyond this point. This is effect is due to diffusion and shift among neighboring sites.

In the implementation of the lattice only immediate neighbors are allowed to interact, but in real life people may interact over relatively large distances in thinly populated areas. This means that the full occupation is a realistic description for the situation in many parts of the world.

\section{Fingerprints of language history seen through qualitatively different datasets}

The data from WALS that were used to produce the examples in Figure 1 involves a mixture of qualitatively different encodings. Features can have from 2 to 9 values and for some there is an internal relationship among these values while for others there is no such relationship.

An example of a binary feature is the presence vs. absence of future as an inflectional category (Dahl and Velupillai 2005). The linguistic typological database developed at the Max Planck Institute for Pscholinguistics in Nijmegen, which was drawn upon by Dunn et al. (2005), consists exclusively of binary features. An example of a WALS feature having nine values is the way that the plural category of nouns is expressed (Dryer 2005). Thus, the plural can be expressed by means of (1) a prefix, (2) a suffix, (3) stem change, (4) tone, (5) a mixture of the preceding, (6) reduplication, (7) a separate word, or (8) a clitic; finally, (9) some languages do not have a nominal plural category.

When there is an internal relationship among the values we shall use the term `ordered feature'. An example of an ordered feature is the inventory of vowel qualities (Maddieson 2005). In Maddieson's formulation of this feature, 2-4 vowels count as `small', 5-6 as `average', and 7-14 as `large'. Thus there are three values. Presumably a language does not change its inventory of vowels directly from small to large but has to pass through a stage where the inventory is average. The same would hold for a change in the opposite direction. This feature, then, is (probably) ordered. The expression of the plural, in contrast, is largely unordered in the sense that a language can mostly change from having any of the possible values to any other. For instance, a language can just as easily go from not having a plural to having either a prefix or a tone expressing the plural. We say ``largely unordered'' because a language may be unlikely, for instance, to go from having no plural to having a mixture of different types of plural (value 5). It is typical for the WALS features that it is not always easy to decide whether they are ordered, semi-ordered or unordered. The feature involving velar nasal consonants (Anderson 2005) may serve as an example of this indeterminacy. Some languages do not have a velar nasal (i.e., the ``ng''-sound of English $thing$), others may have such a sound but not in the beginning of words (as in English), and yet others may allow a velar nasal in the beginning of the word (this is common in, for instance, languages of Africa, South-East Asia or aboriginal Australia). For languages that allow final consonants it is rare to find cases where an initial velar nasal is allowed but not a final one. Thus, if a language goes from not having a velar nasal to allowing velar nasals word-initially we might expect that in most cases it would first pass through a stage where it only had word-final velar nasals (that is, if the language allows final consonants). But this is far from certain and would require more investigation. The example, then, serves to illustrate that classifying WALS features into ordered vs. unordered is an idealization. In reality, order is a matter of degree that would have to be determined by studying how often any pair of values of a given feature co-occur at genealogical language groups with a short history of differentiation. By this method, which has been explored by Michael Cysouw (personal communication, 2006), we could develop an idea about the changes among feature values that are more likely to take place and thus determine to what degree a feature is ordered. To produce Figure 1, the WALS data were treated as unordered.

In the present context we are interested in studying the effects of different numbers of feature values on the outcome of historical linguistic investigations and we also want to know more about the effects of ordered vs. unordered features. For the latter investigation we must necessarily assume that a clean distinction between the two can be made, that is, we assume that the features used are of an ideal type.

For an unordered feature any difference in values will count as 1, whereas for an ordered feature a difference is counted as the absolute difference between the two values. For instance, for the feature of vowel inventory sizes mentioned above, a difference between value 1 (small inventory) and value 3 (large inventory) counts as 2 differences, whereas a difference between value 1 and 2 or value 2 and 3 both count as 1 difference.

\subsection{Features having two vs. more than two values}

\begin{figure}[h]
\includegraphics[scale=0.5,angle=-90]{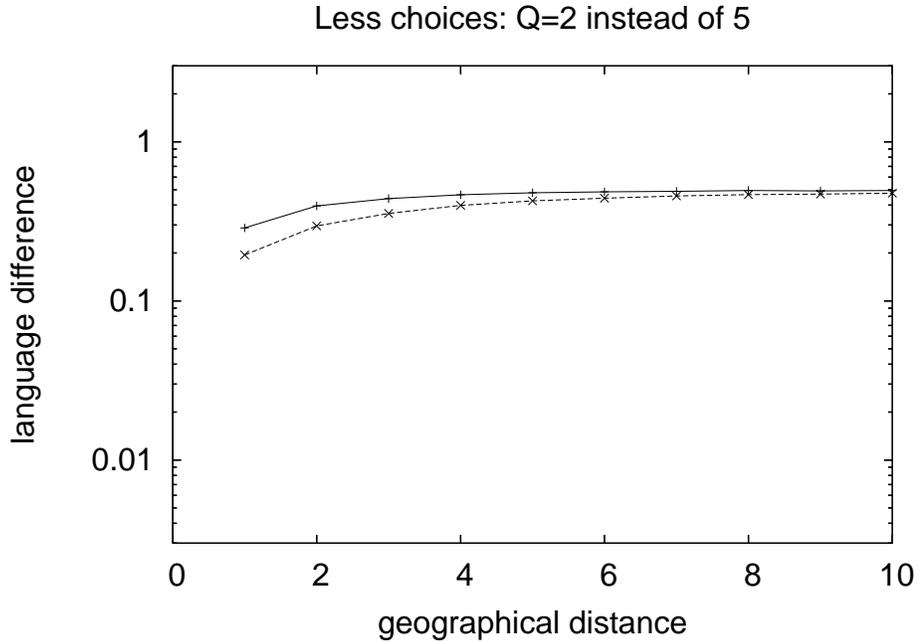}
\caption{The effect of using binary features (here) as opposed to many-valued ones (Figure 2). Legend: + = fragmentation, x = dominance.}
\end{figure}

In Figure 6 we show the differential behavior of binary as opposed to many-valued features. The curves for initial fragmentation and initial domination narrow in on one another for $Q=2$;  this means that it will get harder to discern a difference between related and unrelated languages in binary descriptions.  For $3 \le Q \le 9$ we merely see a parallel shift since the absolute differences may increase for increasing $Q$ as opposed to binary ones. Thus, the curves for $Q=3$ and $Q=9$ are similar to those for $Q=5$ shown in Figure 2, but the curves for $Q=3$ lie lower than the curves for $Q=5$ and those for $Q=9$ lie higher.

\subsection{Ordered vs. unordered features}

Given that ordered features encode more information it is to be expected that such features will preserve the history of the initial difference between dominance and fragmentation better than unordered ones. Figure 7 is nevertheless an important demonstration of the validity of these expected results. It should be compared to Figures 2-4, but in particular to Figure 2, which has the same parameter settings except that the latter involves ordered features.

\begin{figure}[h]
\includegraphics[scale=0.5,angle=-90]{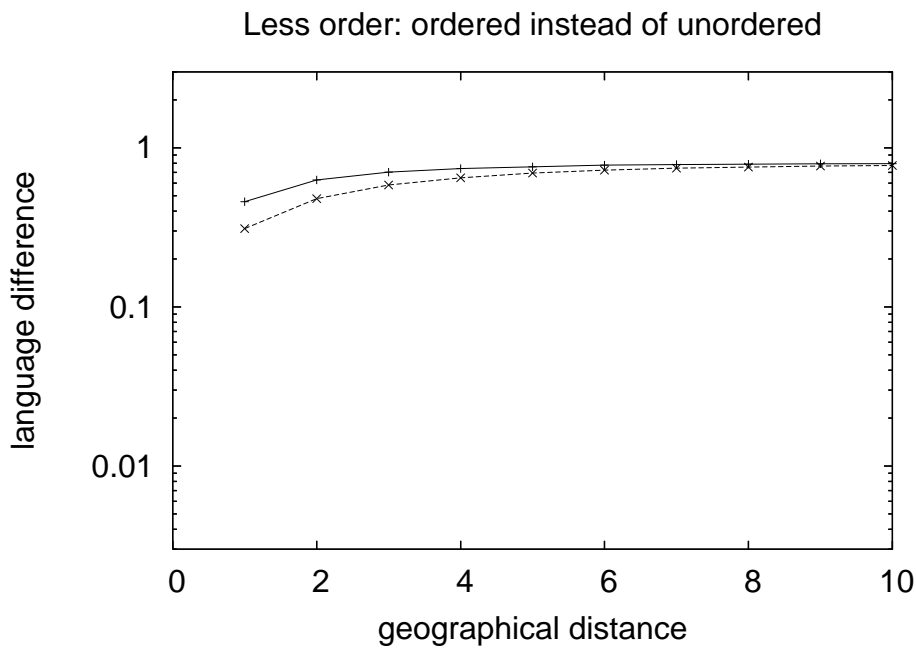}
\caption{The effect of using unordered (this figure) as opposed to ordered features (Figures 2-4). Legend: + = fragmentation, x = dominance.}
\end{figure}

\subsection{Conclusions regarding different results for different encodings of data}

Sections 4.1-2 have shown that the linguist who wants to infer ancient language situations from current distributions of typological data cannot blame it on diffusion, language shift, migration or language change if languages that are known to share a common ancestry nevertheless look equally dissimilar at large geographical distances. Rather, the choice of typological features and the way that they are encoded is expected to be responsible. These two factors, are to a large extent, although not entirely, under the linguist's control. Different typological databases in existence may serve to illustrate the range of choices. The database collected under the auspices of the Pioneers of Island Melanesia project (PIM) of the Max Planck Institute for Psycholinguistics in Nijmegen, i.e. the dataset drawn upon for Dunn et al. (2005), consists entirely of binary features (which are also, by definition, unordered). The database respresented by WALS represents a mixture of everything from binary to 9'ary features, a few of which may be interpreted as ordered, some as unordered, and others as semi-ordered. Finally, the Autotyp database constructed by Balthasar Bickel and Johanna Nichols (http://www.uni-leipzig.de/$\sim$autotyp/) has as many values as the researcher feels necessary for capturing all important differences among functionally related categories. This set of values can be merged to fewer values if such a reduction is opportune for a given purpose. It is also possible to modify the number of feature values in WALS and the PIM data. The former dataset can be recast as binary features, each representing the presence or absence of a given feature value. By the opposite approach, the binary PIM features that pertain to related linguistic categories and have mutually exclusive distributions could be recast as many-valued features. Thus, much is up to the researcher. A linguist cannot decide to make a feature which does not exhibit any ordered behavior ordered. But unordered features can be excluded to strengthen the utility of the dataset for the purpose of making historical inferences. 

\section{General conclusions}

When looking at different parts of the world it is found that typological similarities among languages strongly depend on geography. This dependency is not surprising, but the correlation between typological differences and geographical distance has nevertheless not previously been studied systematically. Known among biologists as isolation by distance, the dependency has been investigated intensively within genetics for well over half a century. While geneticists began emplying computer simulations to study IBD-effects more than a quarter of a century ago, this is the first linguistic study to use such a strategy.

Empirical data drawn from Haspelmath et al. (2005) suggest that the amount of differences among languages may vary for different regions even if the overall IBD pattern is similar. For this reason we set out to investigate how such differences in the curves might come about. Our computer simulations indicate that diffusion may cause languages to become more similar. This is hardly surprising, but a less trivial finding was that the effect is stronger among languages which share a common ancestor than among unrelated languages. Migration apparently contributes little to differences in diversity, although the particular set-up we have chosen for the simulation may be partly responsible for this result. Language shift has a no effect for unrelated languages and only a small effect for related ones. Similarly the effect of language change is only noticeable for related languages. Here, however, the effect is quite drastic. Thus, for related languages diffusion,  language shift, and the rate of internal change may all affect the degree of diversity, whereas unrelated languages should show similar curves for IBD in different parts of the world given that nothing seems to affect these distributions.

Returning to our initial empirical example, we might, then, want to infer that the apparent greater overall similarity at large distances among the languages of the Americas when compared to different parts of the Old World potentially could  be due to descent from a common ancestor or from languages that were similar in structure. If the languages of the Americas were as diverse at the time of the settlement of the continents as the languages of the Old World, we would not expect this difference, since neither rates of language change, language shift, migration or diffusion were found to have much effect on the degree of diversity among languages that start out as being different. Nevertheless, given limitations in the data and the preliminary nature of our computational investigations we do not wish to state this as more than a very weak hypothesis to be investigated in future work. Instead, we wish to focus the readers' attention on the methodological results.

An important finding was that the overall difference in diversity between related (or similar) and unrelated (different) languages is preserved at large geographical distances, at least up to the point where migration and population expansion has resulted in a fully populated area. It is not clear exactly just how fully populated an area must be to be equivalent to the fully populated lattice in our simulations. Thus, it is not certain whether we can expect to find areas in the world that are not fully populated in the sense of the simulations. Nevertheless we expect that for at least some areas it should be possible to verify empirically that related languages may be more similar than unrelated ones at large geographical distances, i.e. that `history is preserved.'

This preservation of history, we also found, is to a great extent dependent on the quality of the data. At large geographical distances, binary or unordered features will obscure the differences between related and unrelated languages, making them look equally dissimilar. This is an important methodological lesson for linguists wishing to make inferences about language history using typological data\textemdash an approach which is becoming increasingly more popular in historical linguistic research.

Computer simulations are a useful tool for making one's assumptions explicit, testing hypotheses, and making predictions about real-world behavior. They do not simply substitute for and illustrate a hypothesis, but represent datasets in their own right. Simulated data need to be interpreted just like ``real'' data. Thus, the final word on the relation between structural diversity and geographical distance among languages is not said in this paper. We expect that other researchers will challenge our interpretations and hope that our findings may be tested by means of alternative simulation models and by attempts to verify or falsify the predictions by means of empirical investigations.

\bigskip
\bigskip
\noindent
\textbf{Correspondence addresses}

\bigskip
\noindent
Eric W. Holman\\
Department of Psychology, University of California, Los Angeles, California 90095-1563, USA. E-mail: Holman@psych.ucla.edu

\bigskip
\noindent
Christian Schulze\\
Institute for Theoretical Physics, Cologne University, D-50923 K\"oln, Germany. E-mail: cs@thp.uni-koeln.de

\bigskip
\noindent
Dietrich Stauffer\\
Institute for Theoretical Physics, Cologne University, D-50923 K\"oln, Germany. E-mail: stauffer@thp.Uni-Koeln.DE

\bigskip
\noindent
S{\o}ren Wichmann (corresponding author)\\
Department of Linguistics, Max Planck Institute for Evolutionary Anthropology, Deutscher Platz 6, D-04103 Leipzig,
Germany (preferred address) \& Languages and Cultures of Indian America (TCIA), PO Box 9515, 2300 RA Leiden, The Netherlands. E-mail: wichmann@eva.mpg.de

\bigskip
\bigskip
\noindent
{\bf \Large Acknowledgements}

\medskip
\noindent
Author names are given in alphabetical order. Holman was responsible for plotting the data in Figure 1, sparking off the inquiry, Wichmann suggested the subsequent research strategy and did most of the write-up, and Stauffer is responsible for the computer simulations, using a model proposed by Schulze. We thank Cecil H. Brown for extensive discussions of many of the issues touched upon here and Brigitte Pakendorf for making us aware of the concept of isolation by distance in biology.

\bigskip
\bigskip
\noindent
{\bf \Large References}

\medskip
\noindent
Abrams, Daniel and Steven H. Strogatz (2003). Modelling the dynamics of language death. \textit{Nature} 424: 900.

\medskip
\noindent
Anderson, Gregory D. S. (2005). The velar nasal ($\eta$). In: Haspelmath et al. (eds.), 42-45.

\medskip
\noindent
Baxter, G. J., R. A. Blythe, W. Crost, and A. J. McKane. (2006). Utterance selection model of language change. \textit{Physical Review E} 73, article no. 046118.

\medskip
\noindent
Cangelosi, A. and D. Parisi (2002). \textit{Simulating the Evolution of Language}. Berlin: Springer-Verlag.

\medskip
\noindent
Christiansen, Morten and Simon Kirby (eds.) (2003). \textit{Language Evolution}. Oxford: Oxford University Press.

\medskip
\noindent
Comrie, Bernard, Matthew S. Dryer, David Gil, and Martin Haspelmath (2005). Introduction. In: Haspelmath et al. (eds.). 1-8.

\medskip
\noindent
Dahl, \"Osten and Viveka Velupillai (2005). Tense and aspect. In: Haspelmath et. al (eds.), 266-281.

\medskip
\noindent
de Boer, Bart (2006). Computer modelling as a tool for understanding language evolution. In: Gontier, Nathalie, Jean Paul van Bendegem, and Diederik Aerts (eds.), \textit{Evolutionary Epistemology, Language and Culture: A Non-Adaptationist Systems Theoretical Approach}, 381-406. Dordrecht: Springer.

\medskip
\noindent
Dryer, Matthew S. (2005). Coding of nominal plurality. In: Haspelmath et al. (eds.), 138-141.

\medskip
\noindent
Dunn, Michael J., Angela Terrill, Geer P. Reesink, Robert A. Foley, and Stephen C. Levinson (2005). Strutural phylogenetics and the reconstruction of ancient language history. \textit{Science} 309: 272-275.

\medskip
\noindent
Epperson, Bryan K. (2003). \textit{Geographical Genetics}. Princeton and Oxford: Princeton University Press.

\medskip
\noindent
Goebl, Hans (1982). \textit{Dialektometrische Studien. Anhand italoromanischer und galloromanischer Sprachmaterialien aus AIS und ALF}. 3 vols. T\"ubingen: Niemeyer. 

\medskip
\noindent
Goebl, Hans (2005). Dialektometrie (Art. 37), in: K\"ohler, Reinhard, Gabriel Altmann, and Rajmund G. Piotrowski (eds.), \textit{Quantitative Linguistik/Quantitative Linguistics. Ein internationales Handbuch/An International Handbook} (Handbücher zur Sprach- und Kommunikationswissenschaft, vol. 27), 498-531. Berlin, New York: de Gruyter.

\medskip
\noindent
Haspelmath, Martin, Matthew Dryer, David Gil, and Bernard Comrie (eds.) (2005). \textit{The World Atlas of Language Structures}. Oxford: Oxford University Press.

\medskip
\noindent
Kosmidis, Kosmas, John M. Halley, and Panos Argyrakis (2005). Language evolution and population dynamics in a system of two interacting species. \textit{Physica A} 353: 595-612. 

\medskip
\noindent
Kretzschmar, William A. (1996). Quantitative areal analysis of dialect features. \textit{Language Variation and Change} 8: 13-39.

\medskip
\noindent
Maddieson, Ian (2005). Vowel quality inventories. In: Haspelmath et al. (eds.), 14-17.

\medskip
\noindent
Mira, J. and A. Paredes (2005). Interlinguistic simulation and language death dynamics. \textit{Europhysics Letters} 69.6: 1031-1034.

\medskip
\noindent
Nerbonne, John, Wilbert Heeringa, Erik van den Hout, Peter van der Kooi, Simone Otten, and Willem van de Vis (1996). Phonetic distance between Dutch dialects. In: Durieux, Gert, Walter Daelemans, and Steven Gillis (eds.), \textit{Proceedings of CLIN '95}, 185-202. Antwerpen.

\medskip
\noindent
Nerbonne, John and Peter Kleiweg (forthcoming). Toward a dialectological yardstick. \textit{Journal of Quantitative Linguistics}.

\medskip
\noindent
Nettle, Daniel (1999a). Using social impact theory to simulate language change. \textit{Lingua} 108: 95-117.
	
\medskip
\noindent
Nettle, Daniel (1999b). Linguistic diversity of the Americas can be reconciled with a recent colonization. \textit{Proceedings of the National Academy of Sciences of the U.S.A.} 96: 3325-3329.

\medskip
\noindent
Nichols, Johanna (1992). \textit{Linguistic Diversity in Space and Time}. Chicago: The University of Chicago Press.

\medskip
\noindent
Niyogi, Partha (2002). The computational study of diachronic linguistics. In: Lightfoot, David (ed.), \textit{Syntactic Effects of Morphological Change}, 351-365. Cambridge: Cambridge University Press.

\medskip
\noindent
Niyogi, Partha (2004). Phase transitions in language evolution. In: Jenkins, Lyle (ed.), \textit{Variation and Universals in Biolinguistics}, xxx-xxx. xxxx: Elsevier Press.

\medskip
\noindent
Niyogi, Partha (2006). \textit{The Computational Nature of Language Learning and Evolution}. Cambridge \& London: The MIT Press.

\medskip
\noindent
Nowak, M. N. Komarova, and Partha Niyogi (2002). Computational and evolutionary aspects of language. \textit{Nature} 417: 611-617.

\medskip
\noindent
Oliveira, Viviane M. de, Marcelo A. F. Gomes, and Ing Ren Tsang (2006a). Theoretical model for the evolution of linguistic diversity. \textit{Physica A} 361: 361-370.
	
\medskip
\noindent
Oliveira, Viviane M. de, Paulo R. A. Campos, Marcelo A. F. Gomes, and Ing Ren Tsang (2006b). Bounded fitness landscapes and the evolution of the linguistic diversity. \textit{Physica A} 368: 257-261.

\medskip
\noindent
Patriarca, Marco and Teemu Lepp\"anen (2004). Modeling language competition. \textit{Physica A} 338: 296-299.

\medskip
\noindent
Pinasco, J. P. and L. Romanelli (2006). Coexistence of languages is possible. \textit{Physica A} 361: 355-360.

\medskip
\noindent
Pr\'evost, Nathalie. 2003. \textit{The physics of language: towards a phase transition of language change}. PhD dissertation, Simon Fraser University.

\medskip
\noindent
Schulze, Christian and Dietrich Stauffer (2005). Monte Carlo simulation of the rise and fall of languages. \textit{International Journal of Modern Physics C} 16: 781-787.

\medskip
\noindent
Schulze, Christian and Dietrich Stauffer (2006). Computer simulation of language competition by physicists. In: Chakrabarti, B. K., A. Chakraborti, and A. Chatterjee (eds.), \textit{Econophysics and Sociophysics: Trends and Perspectives}. Weinheim: WILEY-VCH Verlag.

\medskip
\noindent
Schw\"ammle, V. (2005). Simulation for competition of languages with an ageing sexual population. \textit{International Journal of Modern Physics C} 16.10: 1519-1526.

\medskip
\noindent
S\'eguy, Jean (1973). \textit{Atlas linguistique et ethnographique de la Gascogne}, vol. 6. \textit{Notice explicative}. Paris: Centre national de la recherche scientifique.

\medskip
\noindent
Silva Jr., Wilson A., Sandro L. Bonatto, Adriano J. Holanda, Andrea K. Ribeiro-dos-Santos, Beatriz M. Paix\~{a}o, Gustavo H. Goldman, Kiyoko Abe-Sandes, Luis Rodriguez-Delfin, Marcela Barbosa, Maria Luiza Pa\c c{\'o}-Larson, Maria Luiza Petzl-Erler, Valeria Valente, Sidney E. B. Santos, and Marco A. Zago (2002). Mitochondrial genome diversity of Native Americans supports a single early entry of founder populations into America. \textit{American Journal of Human Genetics} 71: 187-192.

\medskip
\noindent
Sokal, Robert R. and Paolo Menozzi (1982). Spatial autocorrelations of HLA frequencies in Europe support demic diffusion of early farmers. \textit{The American Naturalist} 119: 1-17.

\medskip
\noindent
Stauffer, Dietrich, Suzana Moss de Oliveira, Paulo Murilo C. de Oliveira, Jorge S. S\'a Martins (2006). \textit {Biology, Sociology, Geology by Computational Physicists}. Amsterdam: Elsevier.

\medskip
\noindent
Stone, A. C. and Mark Stoneking (1998). mtDNA analysis of a prehistoric Oneota population: implications for the peopling of the New World. \textit{The American Journal of Human Genetics} 62: 1152-1170.

\medskip
\noindent
Sutherland, William J. (2003). Parallel extinction risk and global distribution of languages and species. \textit{Nature} 423: 276-279.

\medskip
\noindent
Tarazona-Santos, Eduardo and Fabr\'{\i}cio R. Santos (2002). The peopling of the Americas: a second major migration? \textit{The American Journal of Human Genetics} 70: 1377-1380.

\medskip
\noindent
Te\c sileanu, Tiberiu and Hildegard Meyer-Ortmanns (2006). Competition among languages and their Hamming distances. \textit{International Journal of Modern Physcis C} 17: 259-278.

\medskip
\noindent
Tobler, Waldo. 1970. A computer movie simulating urban growth in the Detroit region. \textit{Economic Geography}, 46: 234-240.

\medskip
\noindent
Wang, William S.-Y., J. Y. Ke, and James W. Minett (2004). Computational studies of language evolution. In: Huang, C. R. and W. Lenders (eds.), \textit{Computational Linguistics and Beyond}, 65-106. Institute of Linguistics, Academia Sinica.

\medskip
\noindent
Wang, William S. Y. and James W. Minett (2005). The invasion of language: emergence, change and death. \textit{Trends in Ecology and Evolution} 20.5: 263-296.

\medskip
\noindent
Wichmann, S{\o}ren, Dietrich Stauffer, F. Welington S. Lima, and Christian Schulze. 2006. Modelling linguistic taxonomic dynamics. Submitted to \textit{Transactions of the Philological Society}.

\medskip
\noindent
Wright, Sewall (1943). Isolation by distance. \textit{Genetics} 28: 114-138.

\medskip
\noindent
Zegura, Stephen L., Tatiana M. Karafet, Lev A. Zhivotovsky, and Michael F. Hammer (2004). High-resolution SNPs and microsatellite haplotypes point to a single, recent entry of Native American Y chromosomes into the Americas. \textit{Molecular Biology and Evolution} 21: 164-175.

\end{document}